\documentstyle[eqsecnum,aps,preprint]{revtex}
\newcommand{\en}{{\mathcal N}}
\begin{document}

\tighten

\preprint{
nucl-th/9711066
}
\title{
Thermal Model Analysis of Particle Ratios \\
at GSI Ni--Ni Experiments\\
Using Exact Strangeness Conservation.}
\author{J. Cleymans$^1$, D. Elliott$^1$, A. Ker\"anen$^2$ and E.
Suhonen$^2$}
\address{$^1$Department  of  Physics,  University of Cape Town,
Rondebosch 7701,\\
South Africa\\
$^2$Department of Physical Sciences,\\
University of Oulu, FIN-90571 Oulu, Finland.}
\maketitle

\begin{abstract}
The  production  of hadrons in Ni--Ni at the GSI is considered
in a hadronic gas model with chemical equilibrium. Special attention
is given to the abundance of strange particles which are treated using 
the exact conservation of strangeness.  It is found that all the data 
can be described using a temperature $T =  70\pm 10 $ MeV and a baryon chemical
potential $\mu_B = 720\pm 20$ MeV.
\end{abstract}


\section{Introduction}
Recent experiments at the GSI laboratory have 
shown a relatively large yield of $K^-$ mesons \cite{KAOS,FOPI}
in collisions involving Ni$+$Ni at 1.93 GeV A.
This result is surprising because the beam energy  is 
below the threshold for the reaction $N+N\rightarrow N+N+K^++K^-$
which needs at least 2.5 GeV in the lab. The production of $K^-$ 
mesons therefore proceeds entirely via medium effects in 
the nuclear matter of the colliding nuclei.
An explanation for this phenomenon has been suggested  
in several models \cite{schaffner,brown,weise} 
wherein it is argued that the mass
of the $K^-$ decreases when nuclear density increases while  the
mass  of  the  $ K^+$ increases slightly. 
This would make it easier for
$K^-$  mesons  to  be  produced  since  the  threshold for pair
production becomes considerably lower.
The  production of kaons is
therefore  a  test of the effects of the density of
the medium and should be analyzed carefully.

It  has  been pointed out \cite{PBM,TAPS} that
the  relative   abundance
of hadrons (excluding kaons) in the final state can be
described by using a   hadronic gas
model with chemical equilibrium, i.e.
only  two  parameters, the temperature $T$ and the baryon
chemical  potential  $\mu_B$  describe  all hadronic abundances
except the kaons. 

It  is  the purpose of the present paper to address the 
abundance of kaons in detail. It is well-known that imposing 
exact  strangeness  conservation  introduces  a  suppression of
strange particles if the 
system  has  very  small values of the dimensionless quantity $VT^3$. 
For large values of
$VT^3$  this  suppression disappears rapidly. The size
of  the Ni system  is  relatively  large however the temperatures
involved  are  always  below  100 MeV and thus these corrections
have to be considered seriously.
We  will  show  in  the  present  paper that  all hadronic
ratios, involving strange particles,
 can be explained with only three parameters, namely
the temperature $T\approx 70$ MeV, the baryon chemical 
potential $\mu_B \approx 720$ MeV and
the radius of the system, $R\approx 4$ fm. The large variety of
mesured production ratios of non-strange particles are explained 
with the same set of parameter values.

There  exists  evidence that particle abundances in
heavy ion collisions at higher energies are also very close to chemical 
equilibrium (for a review see e.g. ref.~\cite{sollfrank}). 
The data from CERN favor a region around $T \approx 160 - 200$
MeV and a baryon chemical potential around $\mu_B \approx 180 -
350$  MeV.  The  data  from  BNL favor a region having a lower
temperature,  $T  \approx  100 - 140$ MeV and a larger chemical
potential $\mu_B \approx 450 - 600$ MeV.  The results from GSI 
are in a region which is $T \approx 60 - 100$ MeV and a baryon
chemical  potential  $\mu_B  \approx  700$ MeV. 
The  data  from  GSI thus correspond to a low temperature but a
considerably  higher  value  of  the  baryon chemical potential
$\mu_B$.
Many questions are still left open.
It  doesn't  explain  where  the $K^-$'s are coming from or why
they seem to be so ``easily'' produced despite being sub-threshold.
The  evidence  for  chemical  equilibrium is very strong in our
opinion.  

What  about thermal equilibrium?
 Clearly the momentum
spectra  deviate  strongly  from  a simple thermal distribution
given  by  a  Boltzmann  factor,  however,  many of the effects
cancel  out  when  considering ratios of hadrons as we do here.
Effects due to transverse flow or excluded volume cancel out in
ratios (under circumstances described in \cite{jaipur}).
In  this  paper  we would like to investigate the production of
different particle species as observed at the GSI in Ni--Ni
collisions.  
\section{Formalism}
To   impose  strict strangeness conservation one  projects  the
standard (grand canonical) partition 
function, $Z(T,\lambda_B,\lambda_S,\lambda_Q)$,  onto the state with 
strangeness $S$
\begin{equation}
Z_{S}={1\over 2\pi}\int_0^{2\pi}d\phi\ e^{-iS\phi}
Z(T,\lambda_B,\lambda_S,\lambda_Q) ,
\end{equation}
where the fugacity factor $\lambda_S$ has been replaced by
\begin{equation}
\lambda_S = e^{i\phi}.
\end{equation}
Since  the  incoming state has $S=0$ we will consider only this
value in the rest of this analysis.
For simplicity we present below the formalism used in Boltzmann
approximation.  It  is more complicated to use quantum
`statistics     but     it     can    be    done    (see    e.g.
\cite{mueller,derreth,gorenstein}.
The  partition  function   is then
composed  of  a  sum over the different particle species. Those
that   have   strangeness  zero  are  untouched  by  the  above
operation,  those  that  have  strangeness  +1 acquire a factor
$e^{i\phi}$  while  those  with  strangeness -1 acquire a factor
$e^{-i\phi}$.  If  only  strangeness  $0, \pm 1$ particles are
present we have
\begin{equation}
Z_{S=0}=
{1\over 2\pi}\int_0^{2\pi}d\phi~
\exp \left\{ \en_{S=0}
+\en_{S=1}e^{i\phi}+\en_{S=-1}e^{-i\phi}\right\} ,
\end{equation}
where   $\en_{S=0}$   is  the  number  of  particles  having  zero
strangeness,  i.e.  mainly  pions  and  nucleons.  $\en_{S= \pm 1}$  is
defined  as  the sum over all particles having strangeness $\pm
1$  omitting  the fugacity factor due to strangeness, i.e., the
contribution  of  $\Lambda$  particles  to  $\en_{S=-1}$ would be
given by

\begin{equation}
\en_{S=-1}=V\int   {d^3p\over   (2\pi   )^3}   e^{-  E_{\Lambda}/T
+\mu_B/T} .
\end{equation}

We have checked that the presence of cascade particles does not
modify  our  final results for the analysis of the GSI data. Of
course for higher temperatures cascade particles and $\Omega$'s
should be included.
The strange and anti-strange particle numbers are not
equal since there is a net baryon number in the system, e.g. 
there   is  a difference  in  the  numbers  of  $\Lambda$'s  and
$\bar{\Lambda}$'s  due  to  the  presence of the baryon chemical
potential, $\mu_B$.
\begin{equation}
Z_{S=0}={1\over 2\pi}\int_0^{2\pi}d\phi~
\exp   \left\{     \en_{S=0}
+ \left(\en_{S=1}+\en_{S=-1}\right)\cos\phi
+i\left(\en_{S=1}-\en_{S=-1}\right)\sin\phi \right\} .
\end{equation}
Despite  its  appearance, the above expression has no
imaginary  part.  Exploiting  the  symmetry  properties  of  the
integrand, this can be rewritten as 
\begin{equation}
Z_{S=0}={1\over 2\pi}\int_0^{2\pi}d\phi~
\exp                       \left\{                       \en_{S=0}
+\left(\en_{S=1}+\en_{S=-1}\right)\cos\phi\right\}
\cos\left[\left(\en_{S=1}-\en_{S=-1}\right)\sin\phi
\right] .
\end{equation}
The partition function can thus be written in a compact form as 
\begin{equation}
Z_{S=0}=Z_0{1\over 2\pi}\int_0^{2\pi}d\phi~
\cos\left\{x\sin\phi\right\}e^{y\cos\phi} ,
\end{equation}
where
\begin{equation}
x=\en_{S=1}-\en_{S=-1}
\end{equation}
and 
\begin{equation}
y=\en_{S=1}+\en_{S=-1}
\end{equation}
and  $Z_0$ is the part of the partition function which contains
only non-strange particles.

As an example, the number of kaons is given by 
$$
N_K = gV\int{d^3p\over(2\pi)^3}e^{-E_{K}/T} 
$$
\begin{equation}
\times{{1\over 2\pi}\int_0^{2\pi}d\phi~
\cos\left\{\phi+x\sin\phi\right\}e^{y\cos\phi}
\over
{1\over 2\pi}\int_0^{2\pi}d\phi~
\cos\left\{x\sin\phi\right\}e^{y\cos\phi}} .
\end{equation}
It  is of interest to investigate the small volume limit of the
above  expressions  since  this is where the effects due to the
canonical  formalism  are  most obvious. A typical term has the
following form
\begin{equation}
Z_{S=0} =
{1\over 2\pi}\int_0^{2\pi}d\phi~\exp\left[\cdots
+gV\int{d^3p\over(2\pi)^3}e^{-E/T+i\phi}\cdots\right]  ,
\end{equation}
which in the small volume limit will lead to
\begin{equation}
Z_{S=0}\approx 1 +
{1\over 2\pi}\int_0^{2\pi}d\phi
\left[ gV\int  {d^3p\over (2\pi)^3}e^{-E/T+i\phi}\right]
\times \left[ gV\int  {d^3p\over (2\pi)^3}e^{-E/T-i\phi}\right]
+\cdots .
\end{equation}
Note  that  each  time  there  will  be  two  terms  since  
otherwise the
integration over $\phi$ leads to a vanishing contribution.
The number of $K^+$ will thus be given by
\begin{equation}
N_{K^+}\approx V\int{d^3p\over(2\pi)^3}e^{-E_{K^+}/T}
\left[ gV\int  {d^3p\over (2\pi)^3}e^{-E_{\bar{K}}/T}
 +                       gV\int                      {d^3p\over
 (2\pi)^3}e^{-E_{\Lambda}/T+\mu_B/T}\right].
\label{eq:kplus} 
\end{equation}
This  shows  clearly  that  the  number  of  particles
increase  quadratically  with the volume. For a large system the
dependence on the volume becomes linear.
The  additional  suppression of strangeness can be seen clearly
in   the   small  volume  limit  above.  In  order  to  balance
strangeness,  the production of a kaon has to be accompanied by
either  an anti-kaon or by a strange baryon. This is explicitly
present in equation  \ref{eq:kplus}.

The results are not changed substantially
if  one  includes also particles having strangeness $\pm 2$ and
$\pm 3$ using the analytical methods of \cite{cley1,cley3}.
\section{Numerical Analysis}
\subsection{Resonance Width}
As long as the baryon chemical potential is below $\approx$ 800
MeV  the Boltzmann approximation is adequate. For higher values
one has to include effects due to quantum statistics.
Due  to  the  low temperatures involved the width of resonances
has  to  be  taken  into  account.  For example, an appreciable
number of pions is coming from the decay of $\Delta$ resonances
below  the  mass  of the $\Delta$ but still within the width of
the $\Delta$. 
For  very  small values of the width $\Gamma$, the Breit-Wigner
resonance  shape can be replaced by a $\delta$ function:
\begin{equation}
\lim_{\Gamma\rightarrow 0}{1\over \pi} 
{m\Gamma\over (s-m^2)^2 +m^2\Gamma^2}
 =
\delta(s-m^2)  .
\end{equation}
For large widths this is no longer possible and one has to keep
the  Breit-Wigner  resonance shape. For example the integration
of 
the Boltzmann factor has to be replaced according to 
\begin{equation} \label{eq:BW}
\int d^3 p \exp\left[-\frac{\sqrt{p^2+m^2}}{T}\right]
 \longrightarrow
\int d^3           p           \int    d   s
\exp\left[-\frac{\sqrt{p^2+s}}{T}\right]
\frac{1}{\pi}\frac{m\Gamma}{(s-m^2)^2+m^2\Gamma^2}.
\end{equation}
The mass $\sqrt{s}$ is integrated from $m-2
\Gamma$ to $m+2 \Gamma$. In some cases, 
{\em e.g.} $\Delta(1232), N(1440)$, the
lower limit goes far below the threshold  limit,
e.g.  for the $\Delta$ resonance $\sqrt{s} = m_N + m_\pi$. 
In such cases the lower limit is
chosen to be the threshold value $m_N + m_\pi$.
We  have  taken care of the fact that the normalization must be
adjusted  accordingly  since the integral over the Breit-Wigner
factor should still give unity.
%
%
%
%
%
%
%
%
%
\subsection{Isospin Asymmetry.}
In an isospin asymmetric system there are initially 
four parameters,
namely the temperature $T$, fugacities $\lambda_B$ and $\lambda_Q$, and
the volume $V$. 
The charge chemical potential can be eliminated by considering
the ratio of baryon- and charge content of the system.
The baryon- and charge densities, $n_B$ and $n_Q$, 
and thus the corresponding chemical potentials,
are related  by the condition
\begin{equation}
n_B(T,\lambda,R) = 2\left(\frac{B}{2Q}\right)n_Q(T,\lambda,R),
\end{equation} 
where $(\frac{B}{2Q})$  measures  the isospin asymmetry in the
system. 
For the Ni--Ni system considered here, one has
\begin{equation}
\left(\frac{B}{2Q}\right)_{{\rm Ni-Ni}} = \frac{A_{{\rm Ni}}}{2Z_{{\rm Ni}}}
\simeq 1.04  
\end{equation}
and so there is a 4\% deviation from the isospin symmetric case.
\subsection{Results}
The  hadronic  gas  model  contains all particles listed by the
Particle Data Group \cite{pdg1994}.
Our  main  result  is  shown  in  figure 1. Each 
experimentally measured hadronic ratio corresponds to a band in
the $(T,\mu_B)$ plane. The width of the band corresponds to the
error  bar  which  has  been  reported.  It is to be noted that
several  of  these bands are almost parallel to the temperature
axis.  All  these  bands  are  fairly insensitive to the chosen
hadronic  volume.  The main exception is the $K^+/\pi^+$ ratio.
This  ratio  is  highly sensitive to the value of the radius of
the  hadronic  gas.  In figure 1 we show this for a radius of 4
fm.  The dependence can be seen more quantitatively in figure 2
where  one  can  see the $K^+/\pi^+$ ratio corresponding 
to a radius of 3.7 fm
and  also  for 4 fm. Also shown is the large volume limit which
is  denoted  by  TD  (thermodynamic)  limit,  corresponding  to
infinite
volume.  The  temperature  necessary  to  reproduce  this ratio
rapidly goes down with increasing values of the radius. This is
because  for  a  small volume there is an intrinsic suppression
factor which makes it necessary to go to higher temperature in
order  to reproduce the same ratio. The region of overlap has a
value for the baryon chemical potential of
\begin{equation}
\mu_B = 720 \pm 20~~\mbox{MeV}.
\end{equation}
It  is  not so easy to fix the temperature because many of
the lines are almost parallel to each other, but a temperature interval given by
\begin{equation}
T =  70 \pm 10~~\mbox{MeV}
\end{equation}
will give a good fit to the hadronic ratios.
%
%
%
%
%
%
\subsection{The Ratio $\Lambda/K$}

A requirement of the exact strangeness conservation in an ensemble, where
the hadrons $i$ with $|S_i|\leq 1$ are included, leads to condition
\begin{equation}
        K^+ + K^0 + (K^*) + \overline{\Lambda} + \overline{\Sigma}
        + (\overline{Y}^*)
        = 
        K^- + \overline{K^0} +(\overline{K}^*) + \Lambda + \Sigma + (Y^*),
\end{equation}
where hadron symbols stand for their densities at freeze-out.
Now the baryon chemical potential is quite convincingly fixed to the range
$ 700\ \mbox{ MeV} \leq \mu_B \leq 800\ \mbox{ MeV}$, and the temperature is
found to be around $T \sim 70$ MeV, so the anti-strange baryons can be
neglected due to condition $Y/\overline{Y} \propto \exp(2\mu_B/T)\sim 10^8$.
Further, the heavier resonances (denoted by superscript *) are not
supposed to play any significant role in energies considered, so
\begin{equation} \label{eq:ratio}
        \frac{\Lambda + \Sigma}{K^+ + K^0} \simeq
        1-\frac{K^- + \overline{K^0}}{{K^+ + K^0}}.
\end{equation}  
At the first stage, the effect of isospin asymmetry in the system is neglected,
and we are led to the  result ($\Sigma^0$ and $\Lambda$ are not distinguished) 
\begin{equation}
        \frac{\Lambda}{K^+}  \simeq  1 - \frac{K^-}{K^+} = 0.96
        \pm 0.02,
\end{equation}
where the experimental result $K^+/K^- = 26\pm 9$ is used. 

When the charge chemical potential is taken into account, equation 
(\ref{eq:ratio}) leads to expression
\begin{equation}
        \frac{\Lambda}{K^+} \simeq \frac{(1+\lambda_Q^{-1})
        -(1+\lambda_Q)\frac{K^-}{K^+}}
        {1+\cosh(\ln\lambda_Q)},
\end{equation}
where $\lambda_Q = \exp(\mu_Q/T)$ is a charge fugacity. 
Using values $60\ \mbox{ MeV}\leq T \leq 80\ \mbox{ MeV}$ and
$ 700\ \mbox{ MeV} \geq \mu_B \geq 770\ \mbox{ MeV}$ we extract the 
corresponding values for the charge fugacity (-2 MeV $\geq \mu_Q \geq$ -6 MeV),
 and obtain the result
$\Lambda/K^+ = 0.99\pm 0.03$. Similar procedure gives for neutral kaons the
result $\Lambda/K^0 = 0.94\pm 0.03$.
\section{Summary}
The  data from GSI show very good agreement with a hadronic gas
in chemical equilibrium
having $T \approx 70$ MeV and $\mu_B \approx 720$ MeV.
When comparing the GSI results with those obtained at higher energies
a clear picture emerges. There is an increase in the
 temperature 
 and at the same time a decrease in the baryon chemical potential
 $\mu_B$ in going from GSI to Brookhaven (BNL) and then on to CERN.
This  is indicated schematically in figure \ref{fig3} where we have made
use of the recent compilations of Sollfrank \cite{sollfrank}
(see also \cite{becattini}) and of \cite{TAPS}.
It  is  not  trivial to translate the values for $T$ and $\mu_B$
into  values  for  the energy and the baryon densities. This is
because one needs to know the particel numbers and not just the
ratios  as we have done in this paper. An estimate of these has
been made recently in \cite{greiner}.
A good agreement with chemical equilibrium 
does not mean that the particle spectra should follow exactly a
Boltzmann  distribution  since  the momenta of particles can be
severely  affected by flow. As an example, a model with Bjorken
expansion  in  the longitudinal  direction will still have its
particle  ratios  determined  by Boltzmann factors even though
the  longitudinal  distribution  is nowhere near a Boltzmann
distribution \cite{jaipur}.

The  main  deviation  from chemical equilibrium is observed in
strange   particle abundances.   These   clearly  
deviate  from  chemical
equilibrium  but it seems that a single parameter
measuring the deviation from chemical equilibrium  is sufficient
to  describe  most  of  the  strange  particles.  
The  necessary  suppression can be fully accounted for by using
exact   strangeness  conservation  and  there  is  no  need  to
introduce a strangeness suppression factor $\gamma_S$ as is the
case  for  particle  abundances measured at higher energies at
CERN.
\acknowledgments

We grate\-ful\-ly acknowledge useful and stimulating
discussions with 
R. Averbeck, Peter Braun-Munzinger, V. Metag, 
 Krzysz\-tof Redlich, Dinesh Srivastava and Johanna Stachel. 
 One of
us  (J.C.) gratefully  acknowledges  the  hospitality of the GSI where
this work was started.

\newpage
\begin{figure}
\caption{Curves  in  the
$(\mu_B,T)$ plane corresponding to the  hadronic ratios
indicated.
The interaction volume corresponds to a  radius of 4 fm, and the 
isospin asymmetry
is $B/2Q = 1.04$.}
\label{fig1}
\end{figure}

\begin{figure}
\caption{Curves in the $(\mu_B,T)$ plane showing the dependence
on the radius of the interaction volume.
Interaction volume corresponds to radius 4 fm. 
The $K^+/\pi^+$ and $\phi/K^-$ ratios with the freeze-out radius
$R=3.7$, and the thermodynamic limit for the ratio $K^+/\pi^+$ are
presented with discrete labels. }
\label{fig2}
\end{figure}

\begin{figure}
\caption{Location of the freeze-out temperature and baryon
chemical potential for different energies.}
\label{fig3}
\end{figure}
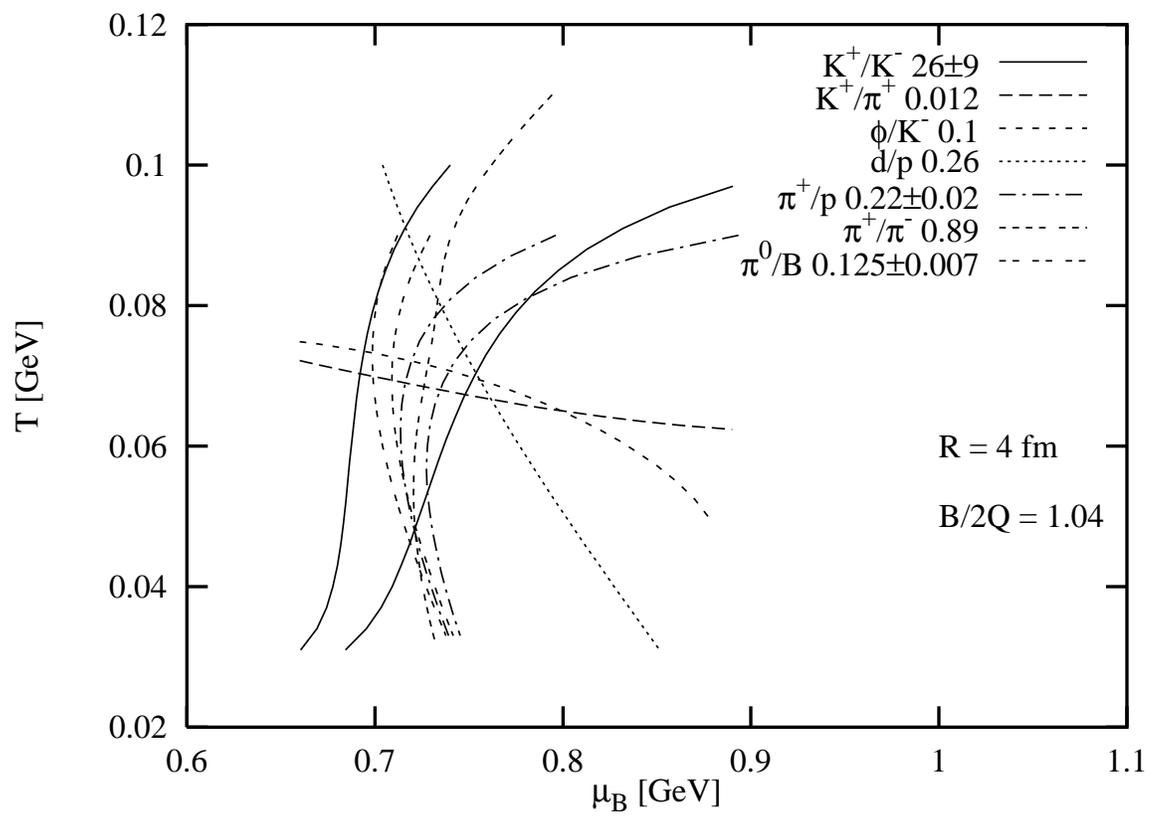
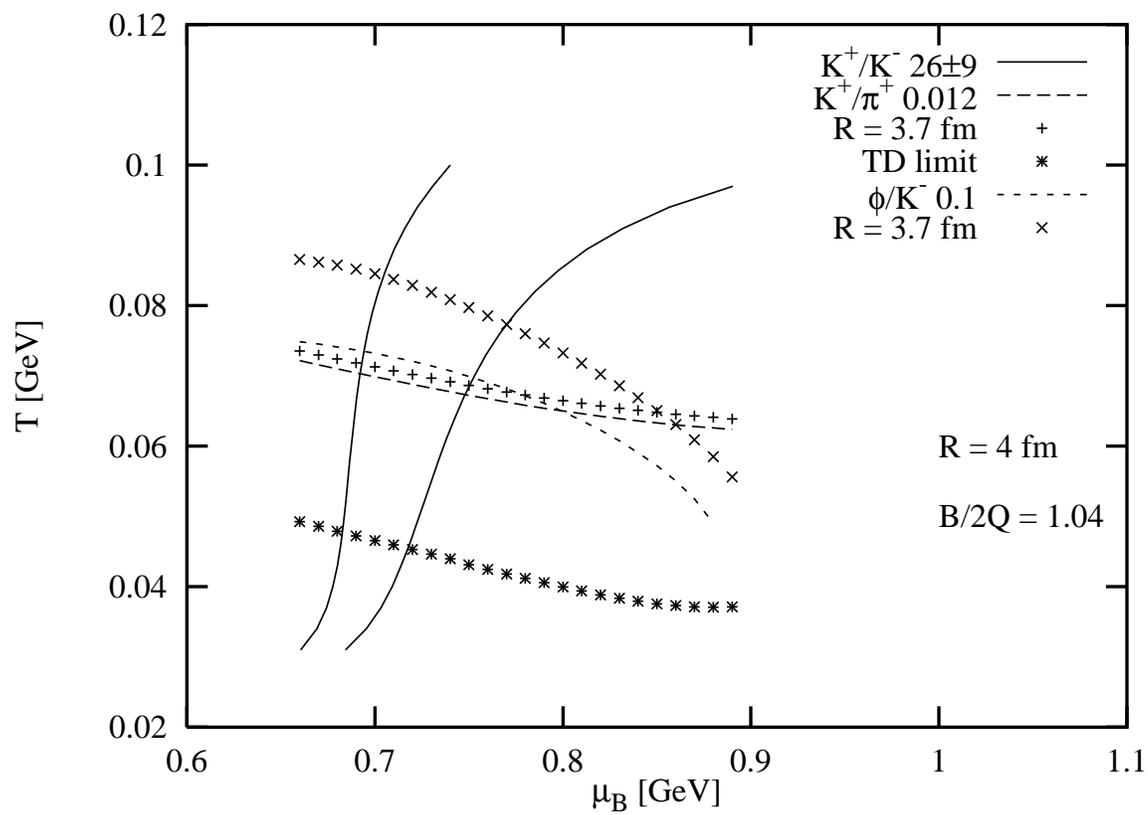
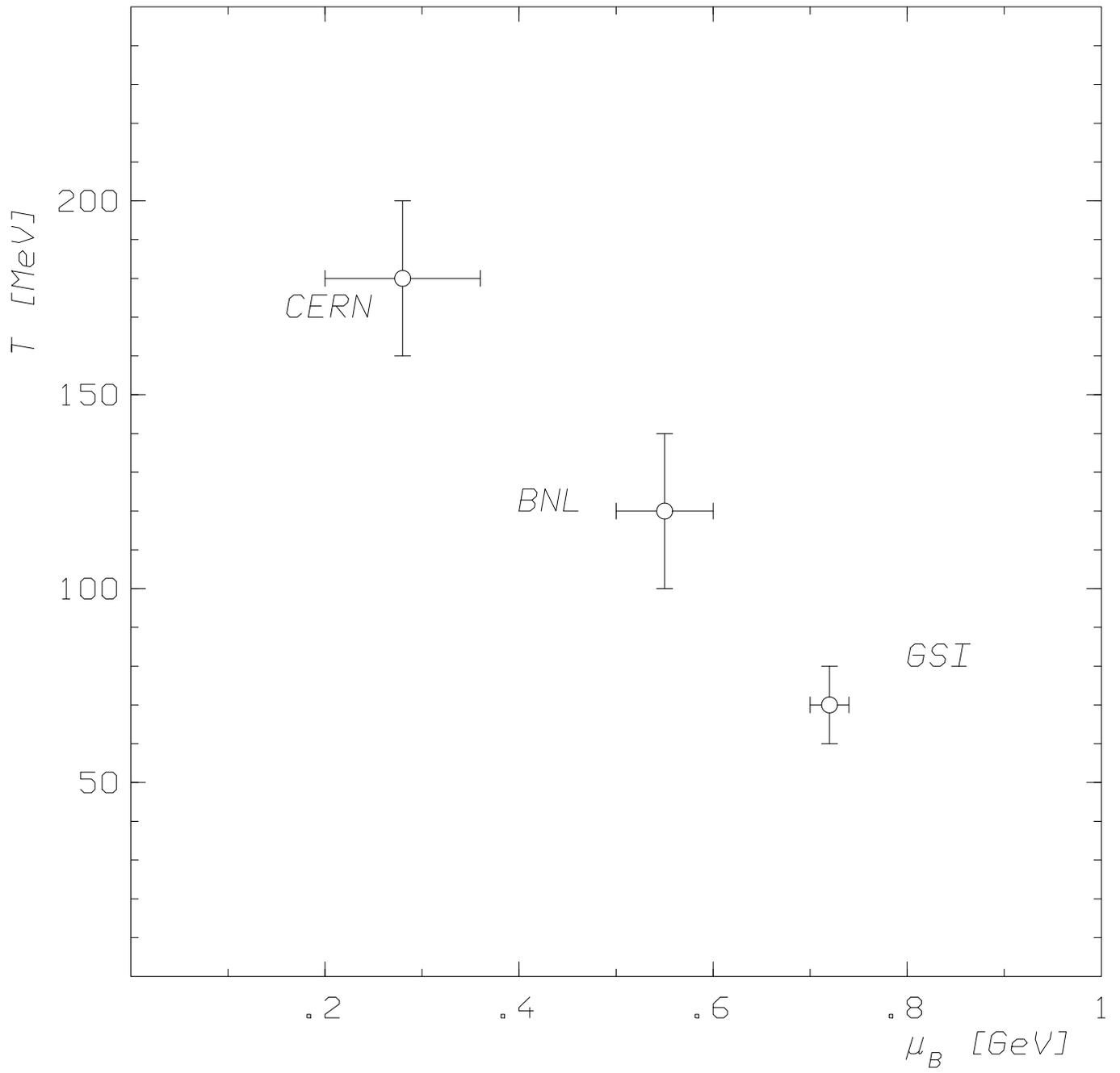
\end{document}